\documentclass[aps,prb,twocolumn,showpacs,amsmath,amssymb,superscriptaddress,floatfix]{revtex4-1}
\usepackage{graphicx}
\usepackage{dcolumn}% Align table columns on decimal point
\usepackage{bm}
\usepackage{epsfig}
\usepackage{epstopdf}
\usepackage{float}	
\usepackage{color}
\include{graphic}

\begin{document}
\title{Electrical detection of magnon-photon interaction via auxiliary spin wave mode}

\author{Peng-Chao~Xu} \email{pcxu14@fudan.edu.cn;}
\affiliation{State Key Laboratory of Surface Physics and Department of Physics, Fudan University, Shanghai 200433, China}
\affiliation{Collaborative Innovation Center of Advanced Microstructures, Fudan University, Shanghai 200433, China}
\affiliation{Department of Physics and Astronomy, University of Manitoba, Winnipeg, Canada R3T 2N2}
\author{J. W.~Rao}
\affiliation{Department of Physics and Astronomy, University of Manitoba, Winnipeg, Canada R3T 2N2}
\author{Y.~Wang}
\affiliation{Department of Physics and Astronomy, University of Delaware, Newark, Delaware 19716, USA}
\author{Y. S.~Gui}
\affiliation{Department of Physics and Astronomy, University of Manitoba, Winnipeg, Canada R3T 2N2}
\author{John Q.~Xiao}
\affiliation{Department of Physics and Astronomy, University of Delaware, Newark, Delaware 19716, USA}
\author{Xiaofeng~Jin}
\affiliation{State Key Laboratory of Surface Physics and Department of Physics, Fudan University, Shanghai 200433, China}
\affiliation{Collaborative Innovation Center of Advanced Microstructures, Fudan University, Shanghai 200433, China}
\author{C.-M.~Hu} \email{Can-Ming.Hu@umanitoba.ca;}
\affiliation{Department of Physics and Astronomy, University of Manitoba, Winnipeg, Canada R3T 2N2}

\begin{abstract}
We report on the electrical detection of a hybrid magnon-photon system, which is comprised of a magnetic sample coupled to a planar cavity. While the uniform Kittel mode has the largest coupling strength among all the magnon modes, it only generates a modest voltage signal by means of inverse spin-Hall effect. We have found that the generated voltage can be significantly enhanced by introducing a higher order magnon mode, which possesses a much higher spin pumping efficiency and furthermore, it is nearly degenerated with the Kittel mode. The experimental results can be explained by our theoretical model, and suggest that the use of an auxiliary magnon mode can realize the configuration of a magnon-photon system with both strong coupling and large spin current.
\end{abstract}

\maketitle

\section{introduction}

Hybrid magnon-photon system has been studied extensively in the last decade as a promising platform for coherent information processing and spintronics devices\cite{Dany2019APE, Soykal2010, Huebl2013PRL, Zhang2014PRL, Tabuchi2014PRL, Goryachev2014PRA, Bai2015PRL, zhang2015NC, Tabuchi2015Science, Yuan2020PRL, Cao2015PRB, Haigh2015PRA, Flaig2016PRB, Bai2017PRL}. The coherent interaction between magnons and cavity photons forms cavity-magnon-polariton (CMP) state, manifests a Rabi-like coupling gap between hybridized mode dispersions\cite{Zhang2014PRL}. While the typical oscillator strength of  magnetic dipole transitions is two orders weaker ($\sim1/137$) than the electric dipole transitions, strong coherent magnon-photon coupling can be also achieved by utilizing the collective spin excitations in a macroscopic ferromagnet\cite{Soykal2010}. Since the magnon-photon coupling strength is determined by the net magnetic dipole moment, it has been found in both theory and experiments that the uniform Kittel mode (UFMR) has the largest coupling strength. \cite{Cao2015PRB, Tabuchi2015Science, Goryachev2014PRA}

Recently, spin pumping in a dynamically coupled magnon-photon system was demonstrated by electrical detection.\cite{Bai2015PRL, Flaig2016PRB} The electrical detection of spin dynamics not only established an alternative method for studying magnon-photon coupling with considerable local tunability, but also created a new avenue to develop the field of cavity spintronics. Successful examples of strong magnon-photon coupling in spintronics include microwave coherent manipulation of pure spin current\cite{Bai2015PRL} and distant spin currents in two YIG/Pt samples\cite{Bai2017PRL}. In spintronic devices, information is carried by spin current and thus a high efficient spin pumping is required in general. Although the UFMR has the largest microwave absorption, it only generates a modest spin current and furthermore, its spin pumping efficiency could be orders of magnitude lower than some higher-order magnon modes.\cite{Sandweg2010APL, Chen2019PRB} Unfortunately, the higher-order magnon modes only weakly couple with the microwave photons since the coupling strength is inversely proportional to the mode number.\cite{Cao2015PRB, Flaig2016PRB} This contradiction motives us to engineer the magnon-photon coupling for pursuing microwave coherent manipulation of pure spin current, where the combination of high spin pumping efficiency and strong magnon-photon coupling is desirable.

In this work, we report on a technique solution in spin pumping based cavity magnonic devices, where the UFMR and a spin wave resonance (SWR, with a finite wave vector) are nearly degenerated and both of them coupled with a common cavity resonance. While the coupling effect caused by the SWR is almost hidden behind the much stronger coupling effect of the UFMR in the microwave transmission measurement, it can significantly boost the overall spin pumping efficiency. Theoretically, we developed a phenomenological model describing the dynamics of the system. The response of all three oscillators show level repulsion with a coupling gap even larger than that formed by each magnon mode and the cavity mode. Experimentally, we have observed spin pumping signal substantially enhanced by the SWR. Our work reveal that the engineering of magnon-photon coupling plays an important role for developing magnonic spintronic devices.

The paper is divided into two main sections, which discuss the theoretical model and experimental results. In the theoretical model part, we construct a coupled magnon-photon system consisting of a UFMR, a SWR and a cavity resonance, which allows us to obtain the dispersions and amplitudes of the hybridized modes. Then we show the significant enhancement of spin pumping current caused by the coupling between SWR and the cavity. Finally, we present the implementation of our experimental setup and quantitatively compare the experimental observations with the theoretical model.

\section{THEORETICAL MODEL}

\begin{figure} [t!]
\begin{center}\
\epsfig{file=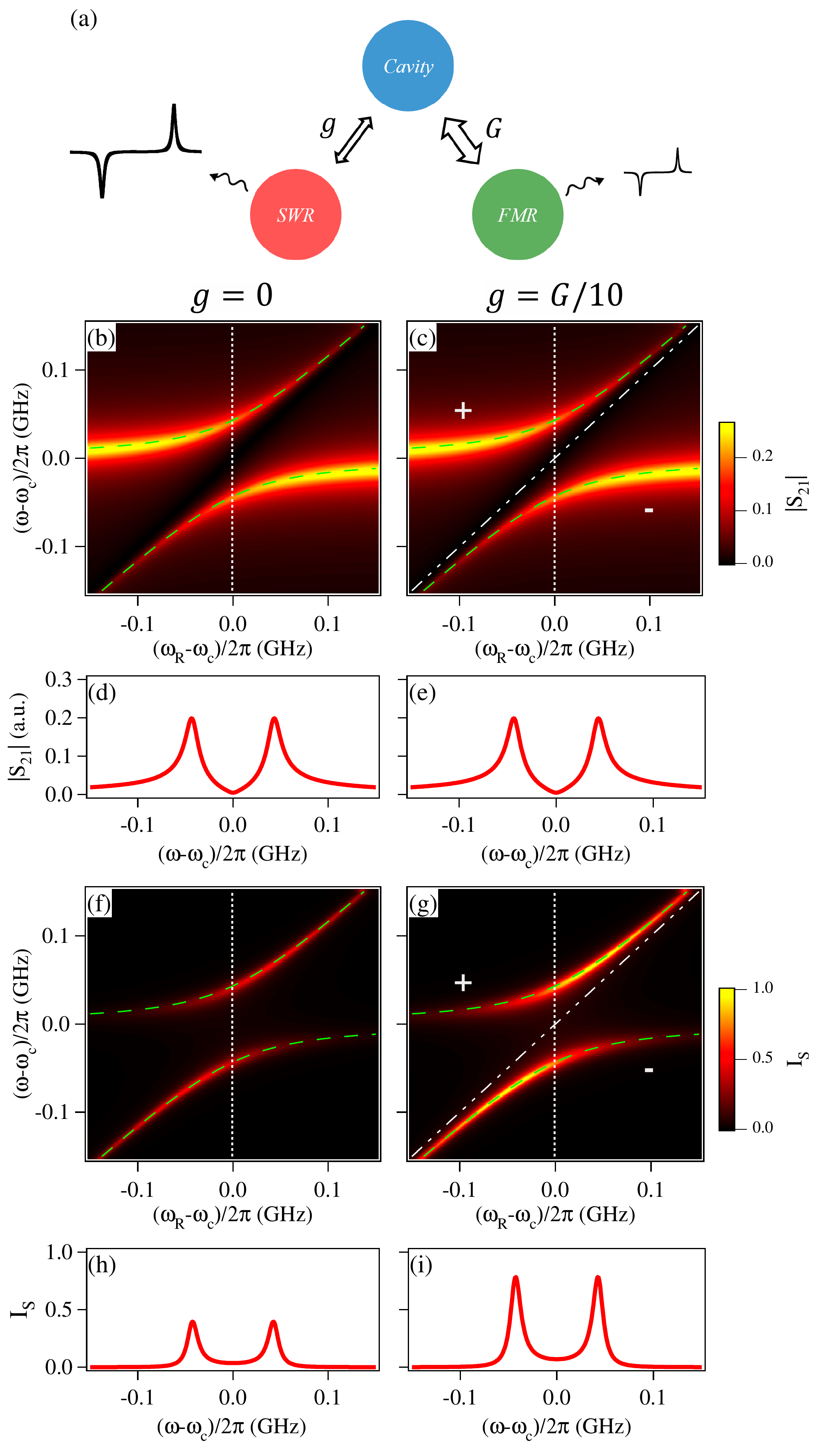, width=8.5 cm}
\caption{(a) A schematic picture of coupled magnon-photon system, in which both UFMR and SWR are strongly coupled to a common cavity mode, but with no direct interaction between them. The calculated microwave transmission are plotted in (b) $g=0$ and (c) $g=G/10$, colored by an identical scale. The calculated total spin current $I_s$ are plotted in (f) $g=0$ and (g) $g=G/10$, colored by an identical scale. (d) (e) (h) and (i) are corresponded spectra at the matched-resonance condition indicated by the white dotted lines in (b) (c) (f) and (g), respectively. For clarity, spin current $I_s$ in (f-i) are normalized by the maximum value in (g). The ratio of the spin pumping efficiency is $\eta_r=100\eta_R$. Other parameters are $\omega_r=\omega_R$, $\gamma_r/2\pi=\gamma_R/2\pi$= 3.5 MHz and $\kappa/2\pi=$2.6 MHz. The in-phase and out-of-phase mode are indicated by $+$ and $-$ in (c) and (g), respectively. Mode dispersions are displayed as green dashed lines and white dotted dashed lines in (b) (c) (f) and (g).
}
\label{Fig1}
\end{center}
\end{figure}

Referring to Fig. 1(a), we study a three-mode coupled magnon-photon system, where an UFMR and a SWR, $M$ and $m$, couple with a common cavity mode $a$. Since different magnon modes in a ferromagnetic thin film in free-space are orthogonal with each other, there is no direct interaction between the UFMR and SWR. The equivalent Hamiltonian of this three-mode system can be written as

\begin{eqnarray}
H=&\hbar \omega_ca^\dag a+\hbar\omega_{R} M^\dag M+\hbar G(a^\dag M+M^\dag a)\\ \nonumber
&\hbar\omega_{r} m^\dag m+\hbar g(a^\dag m+m^\dag a),
\label{Eq:3x3Hamiltonian}
\end{eqnarray}
where $\omega_c$, $\omega_R$ and $\omega_r$ stand for the resonance frequencies of the cavity mode, UFMR and SWR modes, respectively, and $a^\dag$ ($a$), $M^\dag$ ($M$) and $m^\dag$ ($m$) are the corresponding creation (annihilation) operators of them. $G$ and $g$ are the coupling strength of microwave photons with the UFMR and SWR, respectively.

Considering the interaction between the intra-cavity system and photon bath, the quantum Langevin equation can be derived from Eq. (1)

\begin{equation}
\frac{d}{dt}
\begin{pmatrix}
a\\ M\\ m
\end{pmatrix}=
-i\begin{pmatrix}
\widetilde{\omega}_c & G & g\\
G & \widetilde{\omega}_{R} & 0 \\
g & 0 &\widetilde{\omega}_{r}
\end{pmatrix}
\begin{pmatrix}
a\\ M\\ m
\end{pmatrix}+
\begin{pmatrix} \sqrt{\kappa}\\ 0\\ 0\end{pmatrix}P^{in}.
\label{Eq:3x3Langevin}
\end{equation}
Here, $P^{in}$ and $\kappa$ represent the input signal and the coupling rate between photon bath and the cavity mode. $\widetilde{\omega}_c=\omega_c-i(\kappa+\gamma_c)$, $\widetilde{\omega}_R=\omega_R-i\gamma_R$ and $\widetilde{\omega}_r=\omega_r-i\gamma_r$ are the complex frequencies of the cavity, UFMR and SWR modes, respectively. While $\gamma_c$, $\gamma_R$ and $\gamma_r$ are the corresponding intrinsic dampings of them. Since the cavity mode couples with the photon bath with a rate of $\sqrt{\kappa}$, the extrinsic damping term, $\kappa$, appears in the complex frequency $\widetilde{\omega}_c$.

Solving Eq. (\ref{Eq:3x3Langevin}), we obtain the expressions of microwave response

\begin{equation}
\begin{pmatrix} a\\ M\\ m \end{pmatrix}=
\frac{i\sqrt{\kappa}P^{in}}{K}\begin{pmatrix} 1 \\
  \frac{-G}{\omega-\widetilde{\omega}_{R}} \\
  \frac{-g}{\omega-\widetilde{\omega}_{r}}
   \end{pmatrix},
\label{Eq:3x3response}
\end{equation}

\noindent where we define $K=\omega-\widetilde{\omega}_c-\frac{G^2}{\omega-\widetilde{\omega}_{R}}-\frac{g^2}{\omega-\widetilde{\omega}_{r}}$ as the coupling kernel of the system.

The microwave transmission can be determined from Eq. (\ref{Eq:3x3response}) according to
\begin{eqnarray}
  S_{21}&=&\sqrt{\kappa}a/P^{in} \nonumber \\
  &=& \frac{i\kappa}{\omega-\widetilde{\omega}_c-\frac{G^2}{\omega-\widetilde{\omega}_{R}}-\frac{g^2}{\omega-\widetilde{\omega}_{r}}}.
  \label{Eq:S21}
\end{eqnarray}
and the total spin current $I_s$ is the sum of $I_R$ due to the UFMR and $I_r$ due to the SWR, expressed as

\begin{eqnarray}
  I_s&=&I_R+I_r \nonumber\\
       &=&\eta_R|M|^2+\eta_r|m|^2 \nonumber\\
       &=&\eta_R |\frac{\sqrt{\kappa}P^{in}G}{K(\omega-\widetilde{\omega}_{R})}|^2
       +\eta_r|\frac{\sqrt{\kappa}P^{in}g}{K(\omega-\widetilde{\omega}_{r})}|^2.
  \label{Eq:Is}
\end{eqnarray}
Here the parameters $\eta_{R}$ and $\eta_r$ are used to characterize the spin pumping efficiency for the UFMR and SWR, respectively. It is clear that the pure spin current is linearly proportional to the input microwave power (square of the input signal).

An heuristic case occurs when the UFMR and SWR are degenerated and also with an identical damping rate, i.e., $\widetilde{\omega}_R=\widetilde{\omega}_r$. At strong coupling when $G, g\gg\gamma_{R,r,c}$, the eigenfrequencies of the coupled system described above are

\begin{eqnarray}
\omega_\pm&=&\frac{1}{2}\left[\omega_{R}+\omega_c\pm\sqrt{(\omega_{R}-\omega_c)^2+4(G^2+g^2)}\right],  \mathrm{and} \nonumber \\
\omega_3&=&\omega_{r}.
\label{Eq:3x3eigenfrequency}
\end{eqnarray}
The corresponding eigenvectors in ($a$, $M$, $m$) space are [$\omega_\pm-\omega_R$, $iG$, $ig$] for $\omega_\pm$ and (0, $g$, $-G$) for $\omega_3$. The UFMR and SWR precess in-phase for $\omega_\pm$-modes and out-of-phase for $\omega_3$-mode. As a result, $\omega_\pm$  with a coupling strength $\sqrt{G^2+g^2}$ are the bright modes. In contrast, $\omega_3$-mode is a dark mode with net coupling strength $(Gg-gG)/\sqrt{G^2+g^2}$=0, which is completely isolated from the cavity mode so that it cannot be detected from the microwave measurement.\cite{zhang2015NC, Lambert2016PRA, Grigoryan2019PRB, Majer2007Nature, Xu2019PRB} Figures. 1(b)-(e) list calculated transmission spectra of the case $g=0$ and $g/2\pi=0.01G/2\pi=4.3$ MHz. No significant difference can be found from the comparison between the dispersions [Figs. 1(b) and 1(c)] or between the typical spectra at the matched resonance condition [Figs. 1(d) and 1(e)].

However, the spin current of the system are remarkably different when the SWR is only weakly coupled with the microwave cavity. We calculate $I_s$ using Eq. (\ref{Eq:Is}) by assuming $\eta_r=100\eta_R$, which is quite reasonable as we will present in the experiment section. The results are summarized in Figs. 1(f-i). Still, the dark mode is invisible even from magnon dynamics. When compared with microwave measurement, where the signal is stronger near the cavity frequency, the spin current $I_s$ is greater near the frequency of UFMR. More importantly, while the evolution pattern of $I_s$ is identical with or without the SWR, the amplitude of $I_s$ is significantly larger when the SWR is involved. This effect is more clearly seen in Figs. 1(h) and 1(i), where the spectra at the matched-resonance condition are plotted. When compare with the case without the SWR, the maximum spin current is enhanced by $\sim 100\%$. Therefore, by introducing an auxiliary SWR, which has a much larger spin pumping efficiency than the UFMR, we can construct a magnon-photon system with both strong coupling and large spin current.

\begin{figure} [t!]
\begin{center}\
\epsfig{file=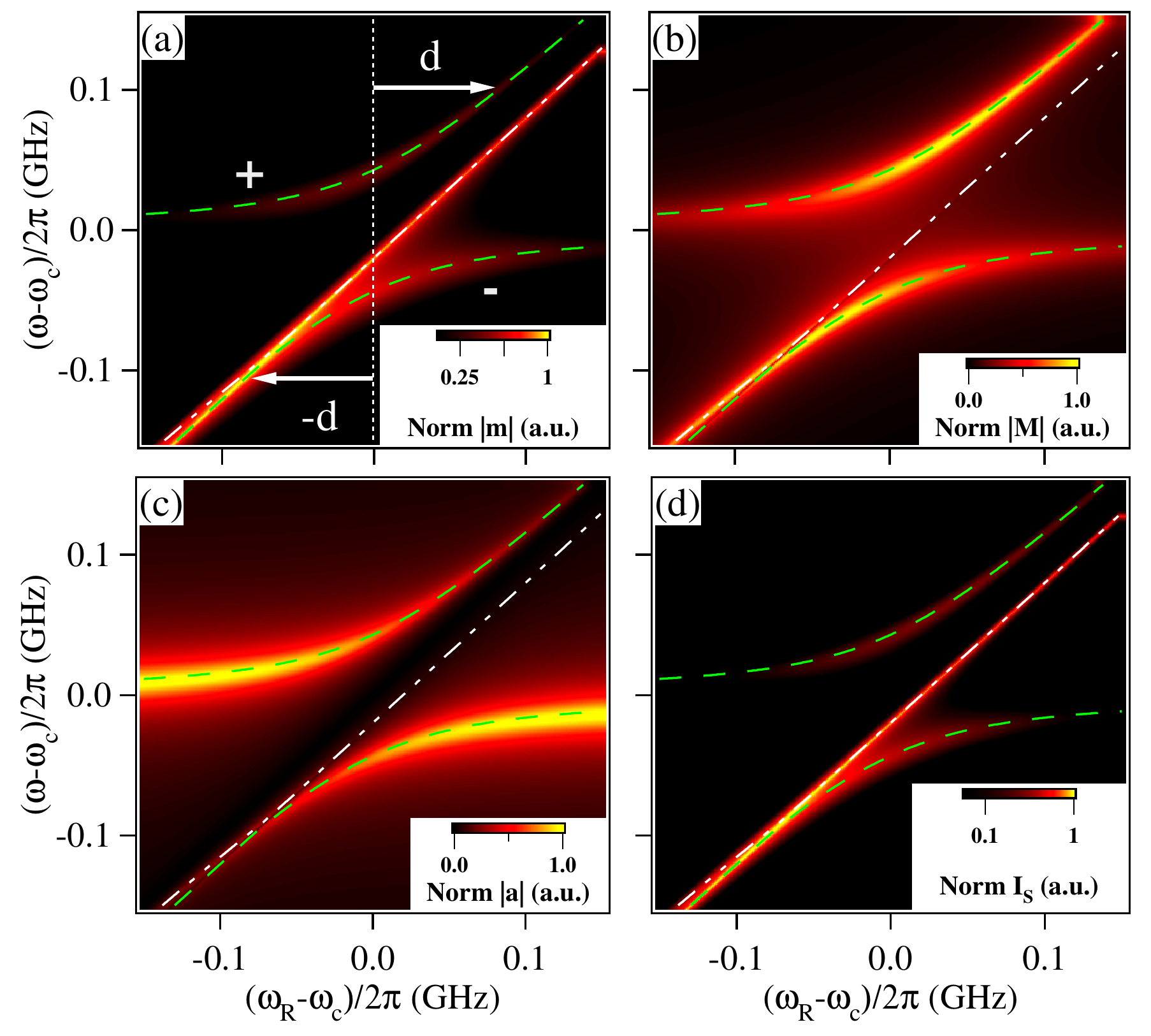, width=8.5 cm}
\caption{(a),(b) Calculated $|m|$ (a) and $|M|$ (b) using Eq. (\ref{Eq:3x3response}), which present diverse magnon dynamics due to the huge difference between coupling rate of $G$ and $g$. (c),(d) Calculated $|S_{21}|$ (c) and $|I_s|$ using Eq. (\ref{Eq:S21}) and Eq. (\ref{Eq:Is}). The white arrows point to $m_\pm$ with opposite detunings. Mode dispersions are displayed as green dashed lines and white dotted dashed lines. All calculations are performed using parameters of our experimental system and assuming $\eta_r=100\eta_R$.}
\label{Fig2}
\end{center}
\end{figure}

In the general case of $\widetilde{\omega}_r=\widetilde{\omega}_R-\delta$, we can solve the corresponding eigenvectors as [1, $iG/(\omega_\pm-\omega_R)$, $ig/(\omega_\pm-\omega_r)$] for $\omega_\pm$ and (-$g\delta$, $iGg$, $-iG^2$) for $\omega_3$. As a result, the $\omega_r$-like mode is no longer the dark mode in principle. Here we calculate the case while setting $\delta/2\pi=20$ MHz, $\gamma_R/2\pi=$ 3.5 MHz, $\gamma_r/2\pi=$1.5 MHz, and focus our study on the case for $g=G/10$ in order to consistent with our experiment. The calculated $|m|$ and $I_s$ are plotted using log scale to show all modes with strong contrast in intensity as shown in Fig. 2.

In this special limit, the $m$ component overwhelms the $a$ and $M$ components in the eigenvector of $\omega_3$. So the contribution of the SWR is dominating in $m$ while negligible in $a$ and $M$ as shown in Fig. 2(a)-(c). A careful examination shows a narrow split along the dispersion of $\omega_3$ as indicated by white dotted dashed lines in (b) and (c). Thus from the response of all parts of the system can we observe dispersions of three modes (indicated by green lines), while the dispersions of $\omega_-$ and $\omega_3$ cross each other due to the frequency difference between the UFMR and the SWR.  

Meanwhile, $|m|$ reaches its maximum as two states overlap, resulting in entirely different intensity evolution $m_\pm$ for $\omega_\pm$ as shown in Fig. 2(a). The asymmetry of mode intensity between $m_\pm$ can be characterized using the ratio of $|m_+|$ and $|m_-|$ with opposite detuning $\omega_R-\omega_c=\pm d$ as indicated by the white arrows. In the case of $\gamma_R\gg \gamma_r$, the ratio can be approximately deduced as 

\begin{equation}
|m_+(d)/m_-(-d)|=|(1-\frac{\delta}{D_-})/(1-\frac{\delta}{D_+})|,
\label{Eq:intensity}
\end{equation}
near the coupling range, where $D_\pm=\omega_R(\pm d)-\omega_\pm (\pm d)$. As far as the dispersions of $\omega_-$ and $\omega_3$ coincide, e.g., $D_-=\delta>0$, $|m_+/m_-|$ reaches zero, indicating substantial amplification of mode intensity of $\omega_-$. The opposite ($D_+=\delta<0$) is true for $\omega_+$. In this sense, we define the special amplification effect as critical amplification. Taking into account both contributions from $m$ and $M$, we calculated normalized $I_s$ using Eq. (\ref{Eq:Is}) and plot it in Fig. 2(d). The mapping replicates almost every feature of $|m|$ owing to the dominating $\eta_r$, including the critical intensity amplification at $D_-=\delta$.

Combine the results of Fig. 1 and Fig. 2, we find that the presence of the auxiliary SWR with large spin pumping efficiency can greatly magnify the spin current signal of the coupled UFMR mode. Moreover, the amplification effect of spin current can be manipulated by the dispersion of SWR dramatically. Specifically speaking, by tuning $\omega_r$, we achieve different frequency regions of $\omega_r=\omega_-$ or $\omega_r=\omega_+$  where the critical amplification effect takes place. In the special case of $\widetilde{\omega}_r=\widetilde{\omega}_R$, the spin current is enlarged by a fixed proportion in the whole frequency range.

\section{Experiment results and discussion}

\begin{figure} [t!]
\begin{center}\
\epsfig{file=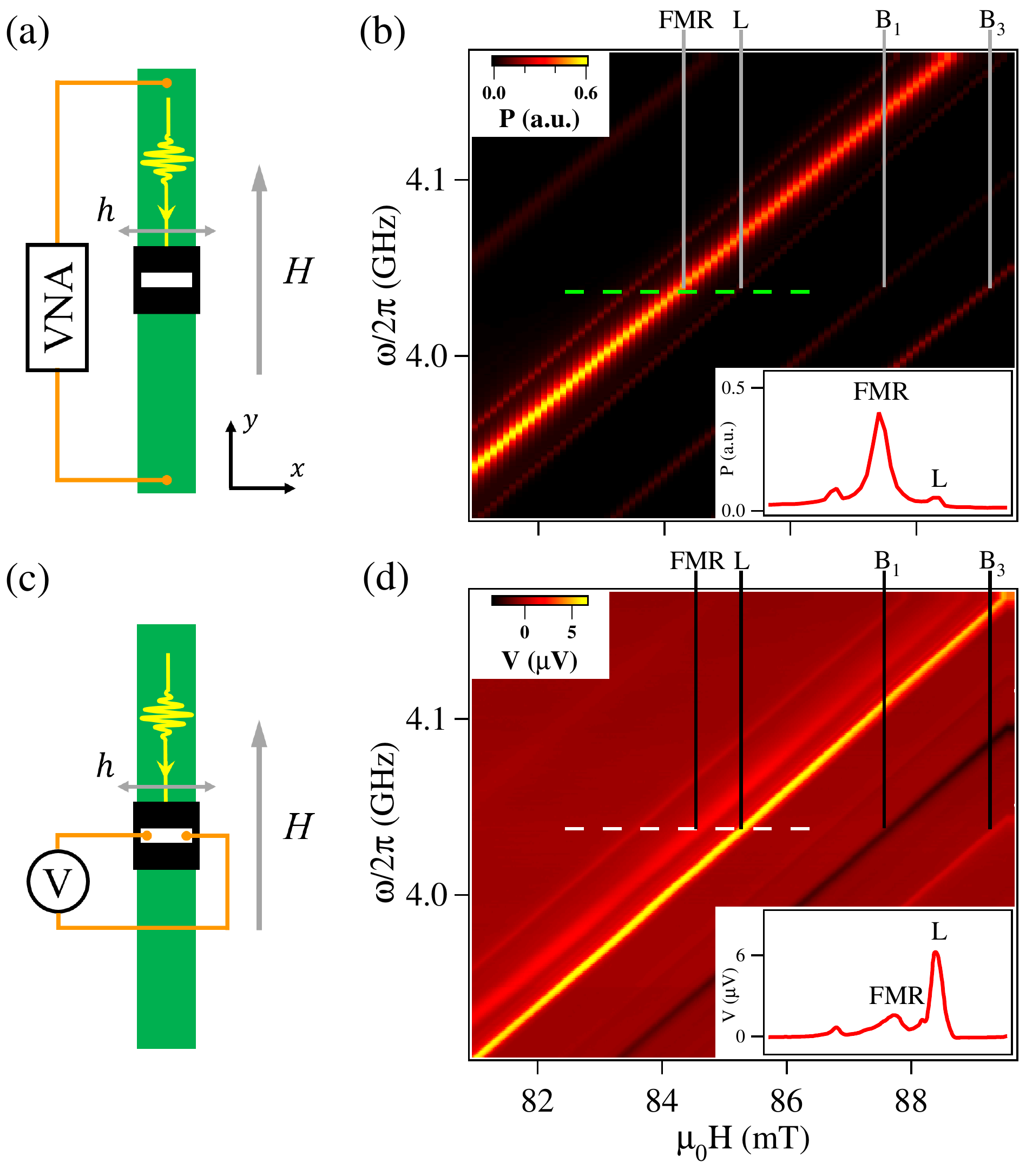,width=8.2 cm}
\caption{(a),(c) Sketch of the experimental setup measuring $S-$ parameter (a) and spin pumping voltage (c) of the uncoupled La:YIG sample. The microwave current carried by the transmission line drives the precessing magnetization through the overlapping of the magnetic insulator and the RF field. (b),(d) Microwave absorption and DC voltage measured while sweeping the magnetic field and the microwave frequency under circumstances of (a) and (c), where the Kittel mode and BVMs are labelled. The insets of (b) and (d) show line cut along the green and white dashed line.}
\label{Fig3}
\end{center}
\end{figure}

Following the proposed approach we implemented a coupled magnon-photon system consisting of a microwave cavity and a ferromagnetic insulator sample. The magnetic sample is 83 $\mu$m-thick La$_{0.03}$Y$_{0.97}$IG (La: yttrium iron garnet) grown on GGG substrate with a dimension of $x\times y$ = 5 mm $\times$ 5 mm. The 5-nm thick Pt layer is patterned into a 3 mm $\times$ 50 $\mu$m stripe for electric detection. Besides the UFMR, the magnon mode used is a higher order magnetostaic (MS) mode\cite{Walker1957PR, Walker1958JAP}. Certain rules are required when selecting suitable MS modes in order to observe a clear critical amplification effect: First, the frequency difference in the dispersions between this mode and UFMR should be much less than $G$, which makes them almost degenerate; Second, the higher order MS mode must produce a dominating spin pumping voltage compare to the UFMR in order to enhance the spin current.

Before studying the coupling effect, we characterize the dispersions and spin pumping voltage of the uncoupled magnon modes. Figure. 3(a) shows the schematic illustration of our experimental setup measuring scattering ($S-$) parameters using a standard 50-$\Omega$ transmission line. In order to generate a homogenous microwave field, the transmission line with a width of 4.8 mm is fabricated on the low-loss Rogers 5880 substrate and the ferromagnetic sample is placed at 5 mm above the geometric center of the transmission line. A tunable external field $H$ is applied perpendicular to the Pt stripe and parallel to the transmission line. The $S-$ parameters can be obtained by connecting both ports of the transmission line to a vector network analyzer (VNA).

Figure 3(b) shows the results of the measured microwave absorption. By sweeping external field $H$ and driving frequency $\omega$, we observe multiple MS modes as potential ingredients of our experiment. The inset of (b) shows microwave absorption along the line cut of the green dashed line. Because of the dominating effective spin number of the Kittel mode, we can identify the peak with highest absorption as the UFMR, which follows the dispersion relation of $\omega_{R}=\gamma\sqrt{|H|(|H|+M_0)}$. Here, $\gamma=27\times2\pi\mu_0$ GHz/T and $\mu_0M_0=0.173$ T are the gyromagnetic ratio and the saturation magnetization of La:YIG sample. Next to the UFMR, higher order backward volume modes (BVMs) \cite{Stancil2009Springer, Zhang2016JAP} are clearly resolved. In addition to the modes $B_i$ with in-plane wave vector $k_i=(k_x,k_y,k_z)=(0,i,0)\pi/d_z$, we also observe BVM $L$ with $k_z = \pi/d_z$, which is about 20 MHz detuned from the UFMR mode and has a microwave absorption of about one-fifth comparing to the UFMR.

The spin pumping measurement was preformed with a similar setup as the $S-$ parameter measurement [Fig. 3(c)]. Alternatively, an amplitude-modulated microwave signal of 23 dBm is feeded into the transmission line. The magnetization precession driven by RF magnetic field produced a nonequilibrium magnetization, which generated a spin current diffusing into adjacent metal\cite{Tserkovnyak2005RMP, Harder2016review}. In the Pt strips, spin current $I_s$ were converted into charge currents via the inverse spin Hall effect (ISHE)\cite{Hirsch1999PRL, Sinova2015RMP, Mosendz2010PRL, Obstbaum2014PRB, Liu2011PRL}, producing spin pumping voltage $V$ that can be measured by a lock-in amplifier. For the $i$th MS mode, we have $V_i\propto I_{si}$.  As clearly seen in Fig. 3(d) and its inset, the BVM $L$ produces a much larger spin pumping voltage (about 4 times larger than the UFMR), which makes it our proposed auxiliary SWR.

\begin{figure} [t!]
\begin{center}\
\epsfig{file=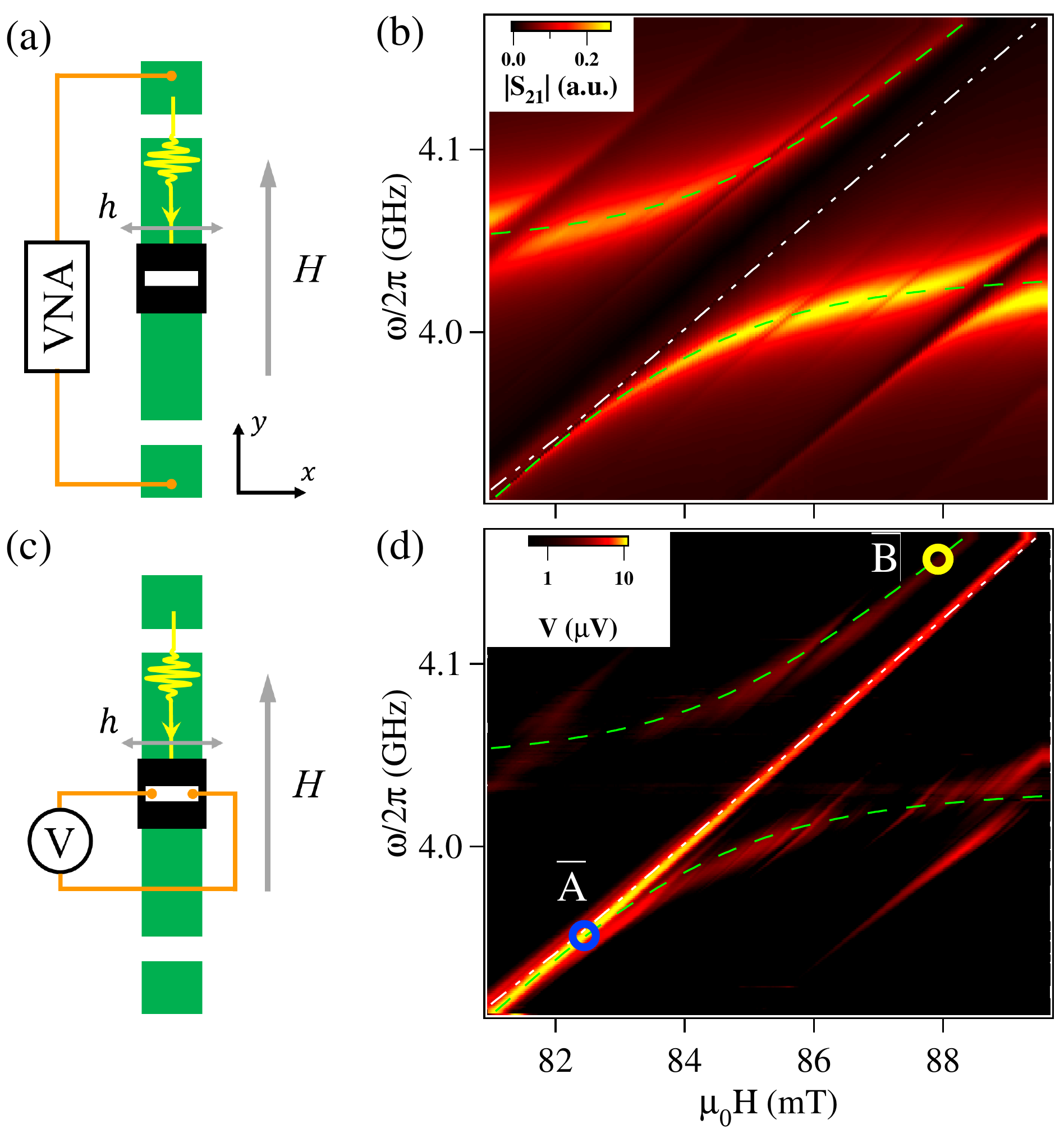,width=8.2 cm}
\caption{(a),(c) Sketch of the experimental setup measuring $S-$ parameter (a) and spin pumping voltage (c) of the La:YIG sample coupled to a stripline resonator. (b),(d) Microwave transmission and DC voltage measured while sweeping the magnetic field and the microwave frequency under circumstances of (a) and (c). The large contrast between voltages of A and B demonstrates the critical amplification effect. Mode dispersions are displayed as green dashed lines and white dotted dashed lines.}
\label{Fig4}
\end{center}
\end{figure}

Next, we couple the ferrimagnetic insulator to a common cavity and measure the transmission spectra and spin pumping signal as illustrated in Figs. 4(a) and (c). Here, the transmission line is replaced by a stripline resonator fabricated on a Rogers 5880 substrate, which has a length of 60 mm, a width of 4.8 mm and a gap of 2 mm to coupled with a standard 50 $\Omega$ transmission line. The resonance frequency of the fundamental mode is $\omega_c/2\pi$ = 4.039 GHz, with a loaded quality ($Q-$) factor of approximate 200.

Figure. 4(b) shows the measured transmission spectra. The main feature of the spectra is a giant avoided crossing, from which we determined the coupling strength $\sqrt{G^2+g^2}/2\pi=43$ MHz. Meanwhile, multiple minor mode splits caused by higher order MS modes are resolved. Particularly, the BVM $L$ manifests a thin dark line on the spectra, cutting through the lower branch of the level repulsion as indicated by the green dashed dotted line. Figure 4(d) presents the measured spin pumping voltage in log scale. By contrast, the mapping shows a dominating signal following the dispersion of mode $L$, sitting in the middle of a shallow signal of level repulsion. To this date, we have determined the essential parameters of our system, which are applied in the calculation results plotted in Fig. 2(d). 

Indeed the experimental observations of electrical detection reproduces three characteristics of the calculation with the estimation of $g=G/10$, $\eta_r=100\eta_R$: (i) The dispersions of all three eigenmodes is presented. (ii) The signal along the dispersion of $\omega_3$ is extremely strong. (iii) The signal reaches maximum (10.7 $\mu$V) when the dispersions of $\omega_-$ and $\omega_r$ coincides at A (indicated by the blue circle). Thus we have demonstrated the enhancement of spin pumping signal via an auxiliary SWR mode. To estimate the maximum magnification of our experiment, we compare the signal of A to the signal of B (indicated by the yellow circle) with opposite detuning $\omega_R-\omega_c$. Since the SWR also contribute to the voltage of B, our estimation implies a lower bound of $V_A/V_B=1650\%$.

\section{Conclusion}

In summary, we have electrically detected spin currents of a La:YIG sample consists of multiple MS modes placed in a planar cavity. Two magnon modes, i.e., the Kittel mode and a backward volume mode are identified and used for configuring a coupling magnon-photon system. Thanks to the high spin pumping efficiency of the auxiliary SWR mode, we observe a significant enhancement of spin pumping signal up to $1650\%$. Moreover, the region of voltage enhancement can be achieved in either whole frequency range or the crossing area of mode dispersions, depending on the frequency of the auxiliary mode with respect to the Kittel model. Our approach reported here provides an alternative way towards engineering pure spin current with vast tunability, hence improves the feasibility and universality of application of electrical detection techniques in hybrid magnon-photon systems.

\section*{Acknowledgements}
This work was funded by NSERC Discovery Grants and NSERC Discovery Accelerator Supplements (C.-M. Hu). John Q. Xiao was supported by DOE, Office of Basic Energy Science under DE-SC--16380. Peng-Chao Xu was supported in part by China Scholarship Council. We thank Yutong Zhao for helpful discussions and suggestions. We also thank Yipu Wang for proofreading the manuscript.

\end{document}